\documentclass[journal]{IEEEtran}
\pdfoutput=1
\usepackage{amsmath,amsfonts}
\usepackage{algorithmic}
\usepackage{algorithm}
\usepackage{array}
\usepackage[caption=false,font=normalsize,labelfont=sf,textfont=sf]{subfig}
\usepackage{textcomp}
\usepackage{stfloats}
\usepackage{url}
\usepackage{verbatim}
\usepackage{graphicx}
\usepackage{cite}

\usepackage{comment}
\usepackage{booktabs}
\usepackage{makecell}
\usepackage{amssymb}
\usepackage[export]{adjustbox}
\usepackage{placeins}
\usepackage{afterpage}
\usepackage{csquotes}
\usepackage{siunitx}
\usepackage{pifont}
\usepackage{amssymb}

\ifCLASSINFOpdf
\DeclareGraphicsExtensions{.pdf}
\fi
\usepackage{subfig}
\usepackage{multirow}
\usepackage{cite}
\usepackage{orcidlink}
\DeclareMathOperator{\sign}{sign}

\newcommand{\anova}[4]{\mbox{$(F(#1,#2)\!=\!#3,~p\!=\!#4)$}}
\newcommand{\anovaPL}[4]{\mbox{$(F(#1,#2)\!=\!#3,~p\!<\!.001)$}}

\newcommand{\ChiSquare}[3]{\mbox{$(\chi^2(#1)\!=\!#2,~p\!=\!#3)$}}
\newcommand{\ChiSquarePL}[3]{\mbox{$(\chi^2(#1)\!=\!#2,~p\!<\!.001)$}}

\newcommand{\wilcoxon}[2]{\mbox{$W\!=\!#1,~p\!=\!#2$}}
\newcommand{\wilcoxonPL}[2]{\mbox{$W\!=\!#1,~p\!<\!.001$}}

\begin{document}
\title{Steering Feedback in Dynamic Driving Simulators: Road-Induced and Non-Road-Induced Harshness}

\author{Maximilian Böhle, \orcidlink{0000-0002-2919-000X} Bernhard Schick, \orcidlink{0000-0001-5567-3913} and Steffen Müller
        % <-this % stops a space
\thanks{Manuscript submitted December 25, 2024}% <-this % stops a space
\thanks{Maximilian Böhle is with the Institute for Driver Assistance and Connected Mobility at Kempten University of Applied Sciences, Kempten, Germany and with the Department of Automotive Engineering at the Technical University of Berlin, Berlin, Germany (e-mail: maximilian.boehle@hs-kempten.de) \\
Bernhard Schick is with the Institute for Driver Assistance and Connected Mobility at Kempten University of Applied Sciences, Kempten, Germany (e-mail: bernhard.schick@hs-kempten.de) \\
Steffen Müller is with the Department of Automotive Engineering at the Technical University of Berlin, Berlin, Germany (e-mail: steffen.mueller@tu-berlin.de)}
}

% The paper headers
\markboth{PREPRINT, SUBMITTED TO THE IEEE TRANSACTIONS ON INTELLIGENT VEHICLES}%
{Boehle \MakeLowercase{\textit{et al.}}: Steering Feedback in Dynamic Driving Simulators: Road-Induced and Non-Road-Induced Harshness}

\IEEEpubid{This work has been submitted to the IEEE for possible publication. Copyright may be transferred without notice, after which this version may no longer be accessible.}

\maketitle

%-------------------------------------------------------------
% Start Paper
%-------------------------------------------------------------

\begin{abstract}
\label{chap_Abstract}
Steering feedback plays a substantial role in the validity of driving simulators for the virtual development of modern vehicles. Established objective steering characteristics typically assess the feedback behavior in the frequency range of up to 30~Hz while factors such as steering wheel and vehicle body vibrations at higher frequencies are mainly approached as comfort issues. This work investigates the influence of steering wheel and vehicle body excitations in the frequency range between 30 and 100~Hz on the subjective evaluation of steering feedback in a dynamic driving simulator. A controlled subject study with 42 participants was performed to compare a reference vehicle with an electrical power steering system to four variants of its virtual representation on a dynamic driving simulator. The effects of road-induced excitations were investigated by comparing a semi-empirical and a physics-based tire model, while the influence of non-road-induced excitations was investigated by implementing engine and wheel orders. The simulator variants were evaluated in comparison to the reference vehicle during closed-loop driving on a country road in a single-blind within-subjects design. The subjective evaluation focused on the perception of road feedback compared to the reference vehicle. The statistical analysis of subjective results shows that there is a strong effect of non-road-induced steering and vehicle body excitations, while the effect of road-induced excitations is considerably less pronounced.
\end{abstract}


\section{Introduction}
\label{chap_Introduction}
\IEEEPARstart{A}{dvances} in modern vehicle development increasingly rely on subjective feedback from drivers in driving simulators. Driver feedback through steering wheel torque (SWT) plays a crucial role in subjective vehicle evaluation \cite{rothhamel2011method,bootz2016fahrwerk,harrer2017steering} and continues to be a research focus in the field of driving simulation. While recent technological progress has allowed substantial aspects of steering development to be addressed in virtual environments, the tuning of steering feel often still involves an iterative process of subjective evaluations conducted by experts in real vehicles \cite{nippold2016analysis,dusterloh2018absicherung,ketzmerick2022validated}. 
With the ongoing transition towards higher automation levels and the increased market focus on Steer-by-Wire systems, conventional development methodologies face new challenges in terms of safety and efficacy. This evolution in the field has led to a growing interest in development methods that enable the transfer of virtual steering characteristics to subjective driver evaluations at any stage of the development process \cite{muenster2014requirements,honisch2015verbesserung,gruener2017objectification}.

\subsection{Related work} 
A substantial amount of previous research has focused on exploring both the role \cite{toffin2003influence,toffin2007role,shyrokau2016influence} and validity \cite{katzourakis2010steering,baumann2014evaluation,gomez2016validation} of steering feel in driving simulators. The decisive role of haptic feedback through SWT in the subjective evaluation of steering feel \cite{barthenheier2007potenzial,wolf2009ergonomische,samiee2015effect} and vehicle dynamics \cite{fiala1967,decker2009beurteilung,gomez2015findings} has been extensively proven. A large number of previous works have highlighted the importance of road feedback for the subjective evaluation of steering feedback \cite{giacomin2004beyond,zschocke2008links,grau2018steering}. While some works on road feedback have pointed out the importance of higher frequency ranges \cite{giacomin2005study,giacomin2006role,berber2009evaluation}, established criteria for the characterization of steering feedback only cover the frequency range of up to \SI{30}{\hertz} \cite{brunn2004objektivierung,groll2006modifizierung,lunkeit2014beitrag}. Although this includes all first-order wheel-induced excitations in typical velocity ranges, the bandwidth of relevant excitations for the subjective evaluation of steering feedback remains the subject of ongoing debate with proposed frequency bands ranging from $30$ to \SI{120}{\hertz} \cite{berber2009evaluation,ajovalasit2013human,grau2016objective}. Nevertheless, investigations into higher frequency ranges of steering wheel and vehicle body vibrations have predominantly approached these as comfort issues, both in terms of human-vehicle interaction \cite{bellmann2002perception,amman2005equal,knauer2010objektivierung} and from the perspective of steering system development \cite{plunt1999strategy,bianchini2005active,an2023active}. 

\IEEEpubidadjcol
The objective of this study is to identify the contribution of road-induced and non-road-induced vibrations between $30$ and \SI{100}{\hertz} on the subjective evaluation of steering feedback in dynamic driving simulators. This covers the transition range between purely tactile and purely acoustic perception, referred to as harshness, which lies beyond the scope of established objective characteristics. This study differs substantially from existing work, both in its methodology and research focus. The present study employs a repeated-measures design to facilitate the direct comparison of driver evaluations obtained from a reference vehicle with those from a dynamic high-fidelity driving simulator. A fundamental prerequisite for this comparability is the physical validity of the evaluated steering representations in the driving simulator and the use of highly accurate road surface models. In electrical power steering systems (EPS), an accurate representation of the EPS software is of paramount importance due to its substantial influence on the steering feedback path \cite{schimpf2016charakterisierung,uselmann2017beitrag,duesterloh2019objectification}. Despite ongoing advancements in the virtualization of steering feedback, the majority of works utilizing high-fidelity steering representations have focused on characterizing steering systems using open-loop testing and calibration work on smooth road surfaces in static simulator environments \cite{wang2016epas,vinattieri2017steering,talarico2021virtual}. Conversely, the prevalence of high-fidelity steering representations in combination with highly accurate road surface models to facilitate the subjective evaluation of steering feedback in dynamic driving simulators is still very limited. 

\cite{shyrokau2015effect,shyrokau2016influence,shyrokau2018effect} and \cite{bertollini1999applying} investigated the effects of steering model complexity and vehicle body motion on the subjective evaluation of steering feel and on-center handling. However, the choice of modifications between both steering feedback and vehicle body motion variants does not allow the application of these findings to the specific effects of particular frequency ranges.  Furthermore, there was no evaluation of road feedback and no representation of road surface in the simulation environment. \cite{parduzi2019method} and \cite{parduzi2021bewertung} investigated the effects of vehicle body motion in the harshness frequency range on driver behavior and performance in a dynamic driving simulator utilizing highly accurate road surface models. However, these investigations did not cover the subjective evaluation of steering feel. Lastly, all presented studies utilized neither steering system representations of sufficient fidelity nor a set of subjective evaluation characteristics that would be suited to address the research questions of this study.

\subsection{Contribution of present research}
This work builds on the findings from our last study in which we showed the beneficial effects of steering wheel excitations between $10$ and \SI{30}{\hertz}, and vehicle body excitations between $10$ and \SI{50}{\hertz} on the subjective evaluation of steering feedback \cite{boehle2024steering}. It extends these findings to the influence of steering wheel and vehicle body accelerations in the harshness frequency range. Furthermore, it enables the isolation of road-induced and non-road induced excitations in this frequency range. The main contributions of this work can be summarized as follows:
\begin{itemize}
    \item Designing and implementing a controlled single-blind subject study to compare the steering feedback in a reference vehicle with four variants of its representation in a high-fidelity driving simulator.
    \item Investigating and isolating the effects of road-induced and non-road-induced steering wheel and vehicle body accelerations between $30$ and \SI{100}{\hertz} on the subjective evaluation of steering feedback in a dynamic driving simulator.
\end{itemize}

\section{Subject study design}
A direct comparison was conducted between the reference vehicle and four variants of its representation in a driving simulator. The sequence of the evaluated simulator variants was randomized between the participants to guarantee the integrity of the single-blind study design and to minimize the impact of familiarization effects. The participants provided written consent for participation, data analysis, and scientific publication. Prior to their driving session, all participants were briefed on the study methodology and evaluation criteria. One lap in the reference vehicle was followed by four laps in the simulator. Both the reference vehicle and the driving simulator variants were rated immediately after each completed lap. Before starting the reference lap, each participant drove approximately six kilometers from the Institute for Driver Assistance and Connected Mobility (IFM) to the reference track’s starting point. This familiarization drive, which took about eight minutes, allowed participants to get accustomed to the reference vehicle. The preparation of measurement data for the driving simulator was performed during the drive back to the IFM. Upon their return, participants proceeded with their simulator drive, if no break was needed. The study was conducted under similar dry weather conditions for all participants.

\subsection{Participants}
An a priori power analysis of the described within-factors repeated-measures design was performed in the program G*Power \cite{faul2007gpower}, version 3.1.9.7. To capture medium sized effects ($f = 0.25$) with a power of $.95$, the minimum sample size was determined to be $31$ with the critical value of \anova{4}{120}{2.447}{.05} . 
In total, $42$ drivers participated in the study. The majority had extensive experience on the driving simulator, otherwise, they were invited to perform a familiarization ride outside the scope of the subject study prior to their participation. Based on their current and past professional experience, the participants were divided into different user levels ranging from $0$ (Normal driver) to $2$ (Steering system expert) according to the definition provided in \cite{boehle2024steering}. Due to the considerable training effects that have been observed in previous simulator studies, particularly during the first hours of simulator usage, all participants without any previous driving simulator experience were assigned the User Level $0$. Two of these participants stated that they were unable to familiarize themselves sufficiently with the driving simulator during the familiarization ride and were therefore unable to perform a full evaluation of all variants. There were, furthermore, three dropouts due to motion sickness and four due to technical issues with data recording during the course of the study, thus reducing the number of valid datasets to $33$. Of these, five participants were female, $28$ male, and the average age was $M = 36.2$ with a standard deviation of $SD = 10.5$ years, covering a range from $24$ to $70$ years in age.  This resulted in a distribution of nine normal drivers, $15$ vehicle experts and nine steering experts. Participation in the study was voluntary and no monetary compensation was provided. Due to the high specificity of the evaluation criteria and the heterogenous subject sample, all participants received an extensive technical briefing prior to the evaluation drives. Furthermore, in line with the findings from previous works on the subjective evaluation of steering feel with non-experts \cite{riedel1997subjektive,krueger1999bewertung,zschocke2009beitrag}, the total study time was limited to one hour and a specifically designed evaluation catalogue was used.

\begin{figure}[!t]
    \centering
    \includegraphics[width=3.5in]{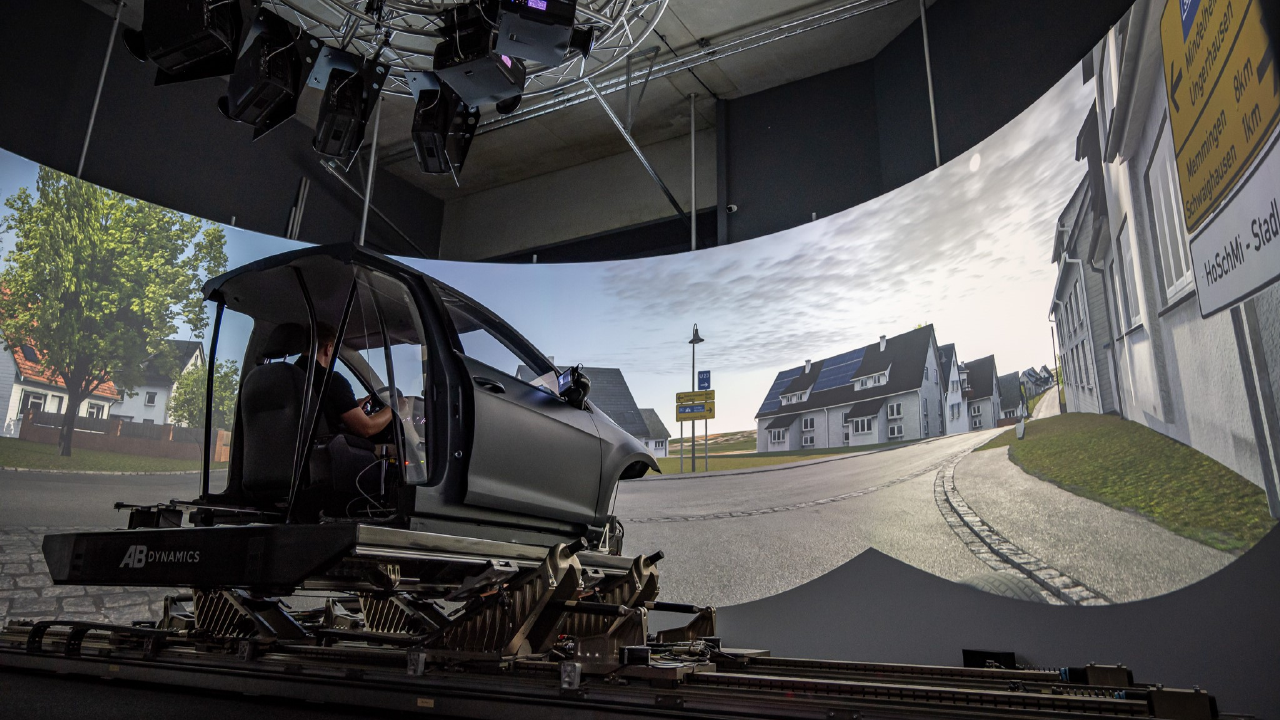}
    \caption{The Advanced Vehicle Driving Simulator
    (aVDS) at the IFM of Kempten University of Applied Sciences on a section of the reference track}
    \label{fig_aVDS}
\end{figure}

\subsection{Reference vehicle}
The reference vehicle utilized in this study was the sports variant of a compact car that featured an EPS system with a progressive gear ratio. During the reference lap, vehicle states were measured using a combined inertial and gyroscopic measurement platform, a GNSS measurement unit with correction data, strain gauge-based force sensors on both front axle tie rods, and a CAN transceiver to capture the vehicle bus signals to and from the EPS ECU. All measurement data were synchronized and resampled to a global sample rate of \SI{1}{\kilo\hertz} through an on-board real-time computer. Prior to their utilization in the driving simulator, post-processing of positions and velocities from GNSS data was performed to compensate for measurement data drift. Prior to conducting the study, a wheel alignment had been performed and except for variations in driver weight, the reference vehicle was kept in constant load conditions throughout the study.

\subsection{Driving simulator}
The driving simulator utilized in this study is the Advanced Vehicle Driving Simulator (aVDS) at the IFM of Kempten University of Applied Sciences, see \autoref{fig_aVDS}. It is equipped with a motion platform driven by eight electric linear actuators, capable of representing vehicle motions with translational accelerations over \SI{10}{\meter/\square\second} and rotational accelerations over \SI{1100}{\degree/\square\second} across all DOFs. The setup includes an EPS Hardware-in-the-loop (HiL) setup with external rack force feedback. The immediate driver environment is represented by a vehicle cabin with a fully functional interior. The visualization system comprises of seven projectors with a refresh rate of \SI{240}{\hertz} on a \SI{270}{\degree} cylindrical screen measuring \SI{8}{\meter} in diameter and \SI{4}{\meter} in height. Vehicle, road, and tire models run on a real-time computer using RedHawk Linux with a global sample time of \SI{1}{\kilo\hertz}. Real-time scheduling of parallel model execution and IO communication via CAN, UDP, and EtherCAT is performed by SIMulation Workbench. A separate real-time computer runs the motion cueing algorithm (MCA) and the controllers for the motion platform at \SI{2}{\kilo\hertz}. Calculation of advanced tire models and EPS HiL communications are oversampled at up to \SI{8}{\kilo\hertz}. Additional information on the physical validity of platform motion and steering feedback can be found in \cite{boehle2024steering}.

\begin{table}[]
    \centering
    \caption{Motion cueing parameters}
      \begin{tabular}{lccc}
      \toprule
      \multicolumn{1}{p{11.285em}}{\textbf{Degree of Freedom}} & \multicolumn{1}{p{2.5em}}{\textbf{Gain}} & \multicolumn{1}{p{5.5em}}{\textbf{High-pass cut-off in Hz}} & \multicolumn{1}{p{5.5em}}{\textbf{Low-pass cut-off in Hz}} \\
      \midrule
      Surge (translation in X)  & 0.5   & 0.50   & 50 \\
      Sway (translation in Y) & 0.5   & 0.25   & 50 \\
      Heave (translation in Z) & 0.5   & 0.50   & 50 \\[0.2cm]
      Roll (rotation around X)  & 0.7   &  -  & 10 \\
      Pitch (rotation around Y) & 0.7   &  -  & 10 \\
      Yaw (rotation around Z) & 0.5   & 0.25   & 50 \\
      \bottomrule
      \end{tabular}%
    \label{tab_MCA}%
  \end{table}%

\subsubsection*{Vehicle motion}
In this study, the aVDS ran a classic washout MCA without tilt coordination, as described in \cite{nahon1985flight} for all translational DOFs and yaw rotation. Roll and pitch rotations were passed to the motion platform directly as scaled angles with an upper frequency limit of \SI{10}{\hertz} without an applied washout. Detailed MCA gain and filter settings are shown in \autoref{tab_MCA}. Motion inputs for all DOFs were limited to a maximum frequency of \SI{50}{\hertz} through the MCA. In some DOFs, the dynamic response of the motion platform imposes constraints on the frequency range that further limit the bandwidth to a maximum between $15$ and \SI{50}{\hertz} \cite{abd2024bandwidth}. The bandwidth of the EPS HiL setup is limited to \SI{30}{\hertz}. Since the scope of the presented study was to investigate influences between $30$ and \SI{100}{\hertz}, the cockpit was equipped with structure-borne sound transducers (SSTs) that are capable of exciting the vehicle body in the entire frequency range of interest. Two SSTs of the type IBEAM VT-200 were used in the vehicle interior, one being placed under the driver seat and one at the inner face of the firewall on the driver side. One SST of the type PUI Audio ASX 11504 was attached to the steering column mounting on the outer face of the firewall. The interior SSTs have an operating range between \SI{25}{\hertz} and \SI{16}{\kilo\hertz}, and the steering SST has an operating range between $5$ and \SI{500}{\hertz}. The SSTs were fed non-road-induced excitations in the frequency range between $30$ and \SI{100}{\hertz}. The influence of acoustic contributions was ruled out using headphones with active noise cancelling.

\subsubsection*{Road and tire model}
The reference road used in this study comprised country road segments with varying surface conditions, resulting in an overall roughness classification of class C. This classification was determined using a single-track evaluation of wavelengths between \SI{10}{\milli\metre} and \SI{10}{\kilo\metre}, as described in \cite{iso8608surface}. The road composition included $19.9\%$ class A, $24.7\%$ class B, and $55.4\%$ class C segments. In the driving simulator, these surface textures were modeled using a horizontal \SI{10}{\milli\metre} grid with a vertical resolution of \SI{1}{\milli\metre}, derived from LiDAR data of the reference road.

In the simulator variants V1 and V2, the tire was modeled as a semi-empirical model implemented as a Magic Formula 5.2 \cite{pacejka2005tire} parameter set. The contact patch was represented using a set of unweighted contact points on the intersection surface with the road surface. Variants V3 and V4 used an FTire \cite{gipser2007ftire} model of the same tire with a physics-based representation of the contact patch and tire belt including tire pressure and temperature effects. The FTire version was r2023.3, the tire wear model was not active. Both tire model datasets were based on the same flat-track dyno measurements and validated regarding their stationary behavior through vehicle measurements. 

\subsubsection*{Vehicle model}
\begin{figure*}[htbp]
    \centering
    \subfloat[]{\includegraphics[width=2.33in]{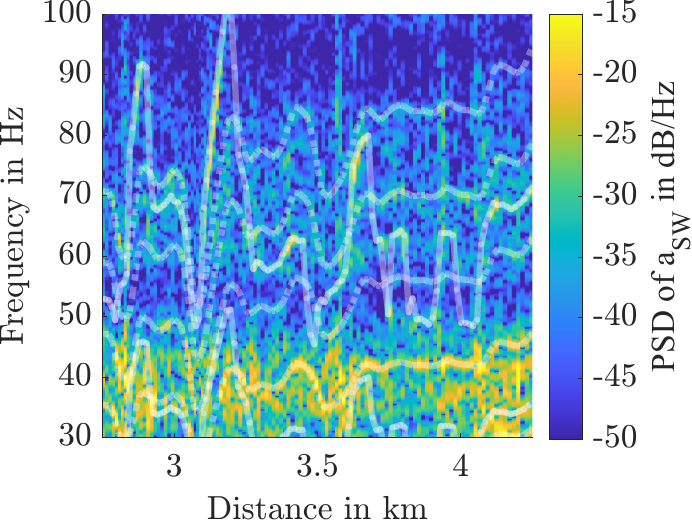}%
    \label{fig_AlgoPerformance_ref}}
    \hfil
    \subfloat[]{\includegraphics[width=2.33in]{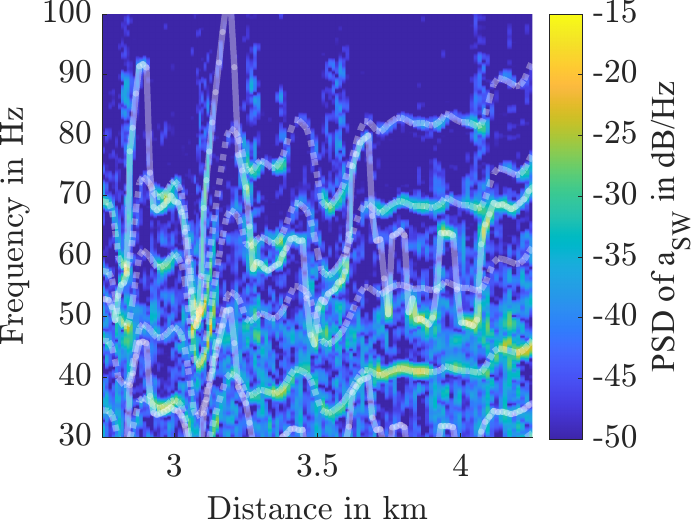}%
    \label{fig_AlgoPerformance_V3}}
    \hfil
    \subfloat[]{\includegraphics[width=2.33in]{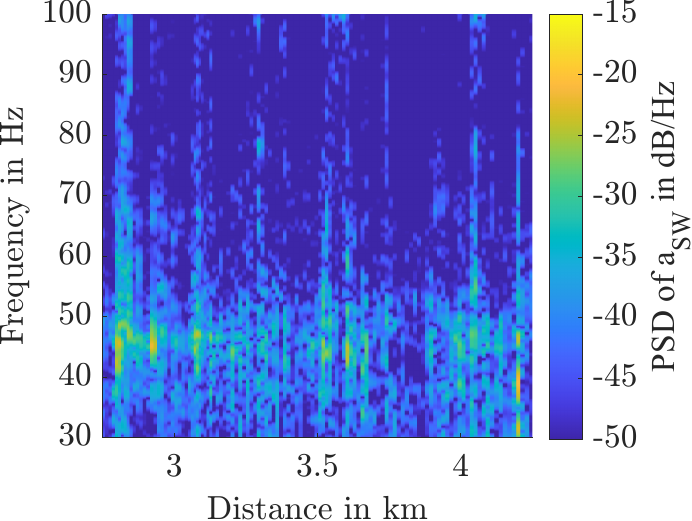}%
    \label{fig_AlgoPerformance_V4}}
    \caption{Spectrograms of measured steering wheel accelerations vs. distance traveled. \ref{fig_AlgoPerformance_ref} Reference vehicle, \ref{fig_AlgoPerformance_V3} aVDS with active SSTs (V1 and V3) and \ref{fig_AlgoPerformance_V4} aVDS with passive SSTs (V2 and V4). The dashed white lines mark the wheel orders while the solid lines mark the engine orders introduced by the SSTs.}
\label{fig_AlgoPerformance}
\end{figure*}

The vehicle was represented by an extended two-track model in IPG Carmaker, treating chassis components and wheels as rigid bodies. Lookup tables that had been parametrized via an elastic ADAMS multibody model and validated on a kinematics and compliance test rig were used to represent elastokinematic effects of the suspension. The validity of the steady-state behavior and transient response of the vehicle model in combination with both tire models had been verified in accordance with \cite{iso22140validation,iso19634validation}. Non-road-induced vibrations were represented through exciter functions based on the vehicle states recorded during the reference drive, such as wheel speeds, engine speed, and torque output. Amplitudes had been validated through IMU measurements of steering wheel accelerations. The resulting excitation profiles are shown in the spectrograms in \autoref{fig_AlgoPerformance}.

\subsubsection*{Longitudinal control}
A critical component of the implemented back-to-back study design was ensuring that drivers experienced the reference track at identical velocities across all compared variants. This was essential for the comparability of objective data in the frequency domain and for validating subjective comparisons. To achieve this, each driver's velocity profile from their reference vehicle drive was recorded. This profile was transformed into a coordinate-based velocity map, which was employed to control the longitudinal motion of the virtual vehicle during all simulator variants, ensuring that the road excitations during the virtual drives matched those during the reference lap. Details on the implementation and performance of the utilized acceleration controller can be found in \cite{boehle2024steering}.

\subsubsection*{Algorithm for rack-force frequency augmentation}
To ensure that only contributions in the harshness range were affecting subjective evaluations, all variants in the presented study made use of the algorithm for Rack Force Frequency Augmentation (RFFA) which had been developed for our previous study \cite{boehle2024steering}. This algorithm applies the bandpassed rack force from the measurements taken during the reference drive to minimize the difference between the vehicle model output and rack forces from the reference vehicle between $10$ and \SI{30}{\hertz}.  

\subsection{Questionnaire}
\begin{figure*}[h!]
    \centering
    \subfloat[]{\includegraphics[width=7in]{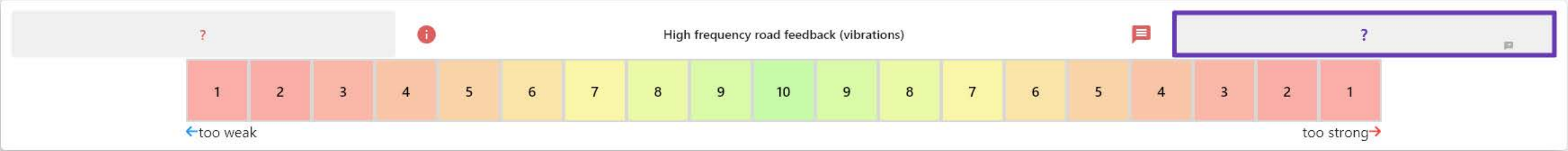}%
    \label{fig_MXeval_BI}}
    \vfil
    \subfloat[]{\includegraphics[width=7in]{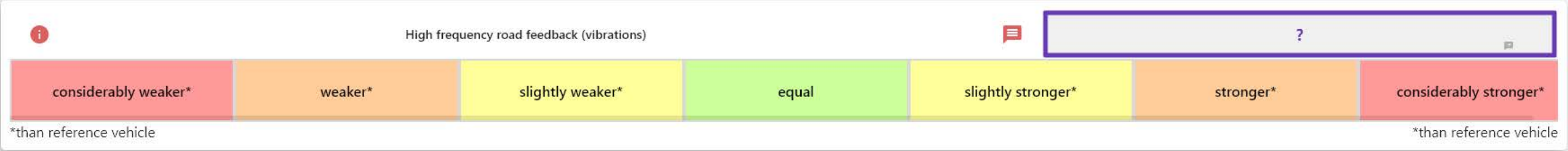}%
    \label{fig_MXeval_Comp}}
    \caption{Screenshot of the questionnaires used for subjective evaluation. \ref{fig_MXeval_BI} BI questionnaire for evaluating system performance and \ref{fig_MXeval_Comp} Comparison questionnaire for evaluating differences between the simulator variants and the reference vehicle.}
\label{fig_MXeval}
\end{figure*}

Post-drive evaluation of the reference vehicle and the simulator variants was performed by means of a questionnaire utilizing the established automotive assessment index (BI) described in \cite{heissing2002subjektive,harrer2007characterisation} to assess system performance in the form of a rating value between $1$ and $10$. A tendency indicating the direction of deviation from the optimum value, e.g. \enquote{too low/high} was reflected in the sign of the BI. Simulator variants were additionally evaluated with a second questionnaire utilizing a seven-point ordinal scale to assess comparability with the reference vehicle in the form of a rating between \enquote{significantly lower} and \enquote{significantly higher,} with the optimum value being \enquote{identical.} Both questionnaires were evaluated using an app, and the user interface is shown in \autoref{fig_MXeval}. The minimum increment size was set to $0.5$, and additional comments could be added via a text box. Similarly to previous studies in this field \cite{zschocke2008links,zschocke2009beitrag,decker2009beurteilung}, this approach aimed to isolate the subjective perception of differences in simulator validity from personal preference.

The evaluation items were based on established criteria for the subjective assessment of steering feedback \cite{riedel1997subjektive,harrer2007characterisation,decker2009beurteilung,rothhamel2011method,gomez2015findings}. A list of these items can be found in \cite{boehle2024steering}. Based on the findings from this work, and in line with the scope of the research questions for this study, the questionnaires contained four additional items. The aim of these additional criteria was to give the participants an opportunity to evaluate the criterion typically referred to as \enquote{Road contact} in more detail. In addition to separating the contributions of steering wheel and vehicle body excitations to road feedback, a distinction was made between individual events that were referred to as Low-frequency excitations (LF) and High-frequency excitations (HF). The additional items were \enquote{Low-frequency steering feedback (bumps and isolated events),} \enquote{High-frequency steering feedback (vibrations),} \enquote{Low-frequency chassis feedback (bumps and isolated events)} and \enquote{High-frequency chassis feedback (vibrations).} Finally, after having successfully completed the study, the participants were asked to choose their preferred simulator variant based on its overall realism.

To summarize, the study employed a 3 (User Level; between-subjects) × 2 (Tire Model; within-subjects) × 2 (SST Status; within-subjects) mixed-factorial design. An overview of the simulator variants is provided in \autoref{tab_variants}.

\begin{table}[htbp]
    \centering
    \caption{Simulator variants: Modification of road-induced excitations using different tire models and non-road-induced excitations using SSTs.}
      \begin{tabular}{ccc}
        \toprule
      \textbf{Variant} & \textbf{Tire Model} & \textbf{SST Status} \\
      \midrule
      V1    & MFTire  & on  \\
      V2    & MFTire  & off \\
      V3    & FTire   & on  \\
      V4    & FTire   & off \\
      \bottomrule
      \end{tabular}%
    \label{tab_variants}%
  \end{table}%

\section{Results}
Due to the heterogeneity of subjective data regarding their normality and homogeneity of variances, the analysis of subjective data was divided into two steps. Initially, the central tendencies of raw evaluation data were analyzed to obtain an overall ranking of the evaluated variants. After analyzing the result distributions, data were transformed for subsequent statistical analysis of the expected main effects. In addition to hypothesis testing, post-hoc comparisons of selected variants were performed. In accordance with the questionnaire structure, the data analysis was divided into the subjective assessment of system performance by means of the BI and the comparison of the realism of simulator variants through the comparative ordinal scale. 

\subsection{Distribution of raw subjective data}
\begin{figure*}[!t]
\captionsetup[subfigure]{labelformat=empty}
\centering
\subfloat[V1]{\includegraphics[width=2.8in]{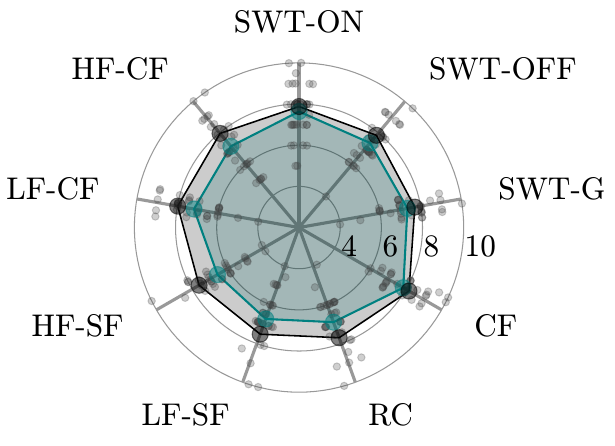}%
\label{fig_Spider_RoadContactVar1_311}}
\hfil
\subfloat[V2]{\includegraphics[width=2.8in]{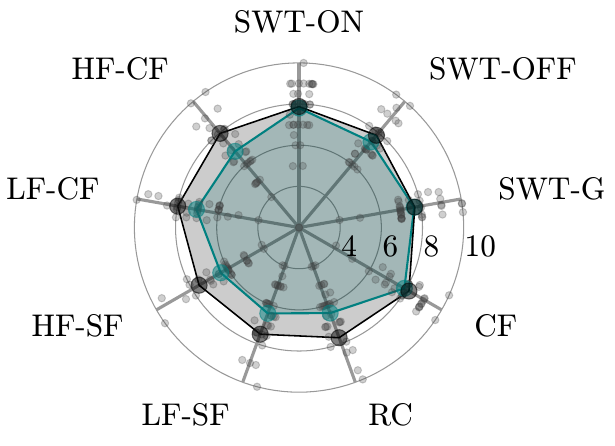}%
\label{fig_Spider_RoadContactVar2_311}}
\vfil
\subfloat[V3]{\includegraphics[width=2.8in]{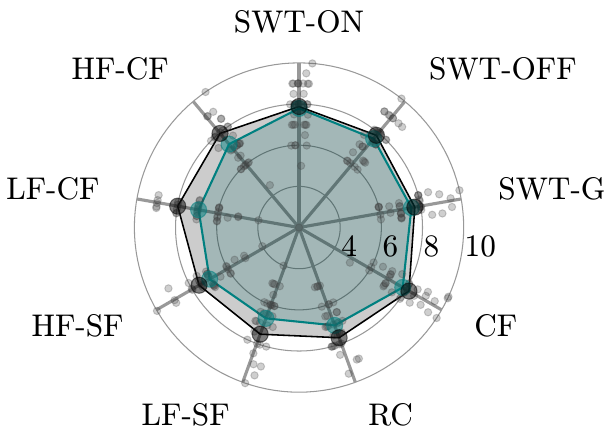}%
\label{fig_Spider_RoadContactVar3_311}}
\hfil
\subfloat[V4]{\includegraphics[width=2.8in]{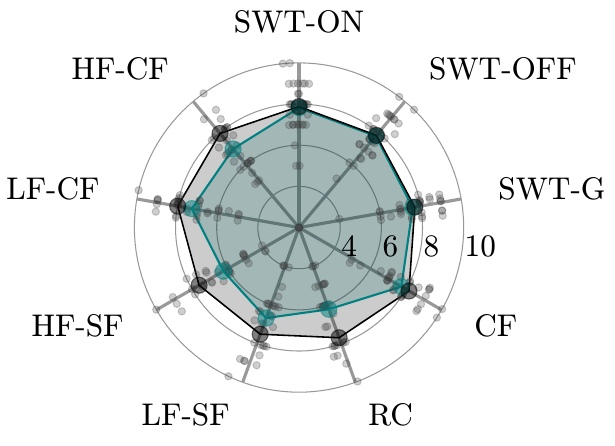}%
\label{fig_Spider_RoadContactVar4_311}}
\caption{Absolute values of BI ratings (higher value means better). Only the value range between $4$ and $10$ is shown. The mint-colored lines and surfaces represent the simulator variants while the black lines represent the reference vehicle. SWT-ON: \enquote{SWT on-center,} SWT-OFF: \enquote{SWT off-center,} SWT-G: \enquote{SWT gradient,} CF: \enquote{Center feel,} RC: \enquote{Road contact,} LF-SF: \enquote{Low-frequency Steering feedback,} HF-SF: \enquote{High-frequency Steering feedback,} LF-CF: \enquote{Low-frequency Chassis feedback,} HF-CF: \enquote{High-frequency Chassis feedback}}
\label{fig_Spider_subFigures}
\end{figure*}

An initial inspection of absolute values of subjective ratings revealed clear tendencies regarding both the general rank of evaluated variants and the effect direction of the influences under investigation. Ratings of stationary characteristics show smaller differences than those related to Road contact. The reference vehicle outperformed all simulator variants in all evaluated characteristics. Variants V1 and V3 with active SSTs were ranked better and equally or more realistic in all criteria related to Road contact compared to variants V2 and V4 without active SSTs. V3 with the FTire model and active SSTs ranked most realistic in all criteria related to Road contact, and best and most realistic in both criteria related to the representation of high frequencies. V3 was rated as the most realistic variant both in most individual criteria and by most participants in terms of overall realism. V1 was rated as the least realistic in most individual criteria. \autoref{fig_Spider_subFigures} shows these results in comparison to the reference vehicle. 

\subsection{Data preparation for statistical analysis}
Individual differences in scale usage between the participants lead to considerable heterogeneity in terms of variance and normality. This variability is expected both on an individual level and between different User Levels, since each participant applies a unique distribution to the BI scale, both with regard to its central tendency and its spread. The mean value differences between all simulator variants and the dependence of the evaluation variance on the User Level is shown in \autoref{tab_MeanVar_Variant_UserLevel}. While all user groups show some degree of variance in scale usage, overall result variances are considerably lower for higher User Levels. For this reason, a subgroup analysis of the participants with a higher User Level can be beneficial despite decreased statistical power due to the reduced sample size.

\begin{table*}[htbp]
    \centering
    \caption{Mean values of absolute BI evaluations (Higher means better) split by variant and variances split by User Level.}
      \begin{tabular}{lcccccccccc}
      \toprule
      \textbf{ }  & \textbf{Variant}  & \multicolumn{3}{c}{\textbf{SWT}}                              & \textbf{Center} & \textbf{Road}     & \multicolumn{2}{c}{\textbf{Steering Feedback}}  & \multicolumn{2}{c}{\textbf{Chassis Feedback}}  \\
                  &                   & \textbf{On-Center}  & \textbf{Off-Center} & \textbf{Gradient} & \textbf{Feel}   & \textbf{Contact}  & \textbf{LF} & \textbf{HF}                       & \textbf{LF} & \textbf{HF}                      \\
        %\specialrule{0.1pt}{2\jot}{1pc}
        \midrule
      \textbf{Mean}         &   Ref	&	7.88	&	7.85	&	7.71	&	8.17	&	7.70	&	7.52	&	7.61 & 7.98 & 7.97    \\
                            &   V1	&	7.64	&	7.36	&	7.35	&	7.85	&	6.89	&	6.73	&	6.58 & 7.18 & 7.15   \\
                            &   V2	&	7.80  &	7.44	&	7.68	&	7.91	&	6.42	&	6.44	&	6.39 & 7.06 & 6.83   \\
                            &   V3	&	7.79	&	7.64	&	7.55	&	7.82	&	7.06	&	6.70	&	7.02 & 6.95 & 7.29   \\
                            &   V4	&	7.82	&	7.80  &	7.61	&	7.74	&	6.23	&	6.68	&	6.24 & 7.29 & 6.98   \\[0.2cm]
      \textbf{ } & \textbf{User Level} & \textbf{} & \textbf{} & \textbf{} & \textbf{} & \textbf{} & \textbf{} & \textbf{} \\
                            \midrule
        \textbf{Variance}   &   0     & 2.23	&	1.66	&	1.79	&	2.20	&	2.80	&	3.44	&	3.40 & 1.90 & 1.71    \\
                            &   1     & 1.05	&	1.39	&	1.22	&	1.13	&	1.75	&	2.03	&	1.49 & 1.23 & 1.74    \\
                            &   2     & 0.91	&	1.12	&	1.12	&	1.07	&	2.14	&	0.691	&	1.64 & 1.69 & 1.99    \\
      \bottomrule
      \end{tabular}%
    \label{tab_MeanVar_Variant_UserLevel}%
  \end{table*}%

A further contribution to the non-normality of raw subjective data lies within the orientation of the bilinear BI scales used for the majority of evaluation criteria, with the expected values close to either end of the scale instead of its center. To address the discontinuity in the evaluation results, the sign-weighted BI was transformed into its sign-weighted deviation from the ideal BI rating of ten, preserving the directional information conveyed by the sign. The implementation of this transformation is shown in \autoref{eq_inversion}.

\begin{equation}
  \label{eq_inversion}
  BI_{inv} = \sign(BI) \times (10 - |BI|)
\end{equation}

The resulting scale has its center at zero, with a positive deviation representing an item rated as \enquote{too high/strong/steep,} while a negative deviation represents an item rated as \enquote{too low/weak/flat.} Subsequently, all ratings from the first questionnaire were normalized using a z-transform to further mitigate the effect of different scale usage by individual users or user groups. As shown in \autoref{eq_zTransform}, the z-transform returns a $z$-score for each subjective rating. It is applied across all criteria for each user to normalize their responses relative to their own response distribution. This results in a centered, scaled distribution with a mean value of zero and and a standard deviation of one.

\begin{equation}
  \label{eq_zTransform}
  z = \frac{(x - \overline{X})}{S}
\end{equation}

wherein $x$ is the current value of the input sample, $\overline{X}$ is its mean value, and $S$ represents its standard deviation. 

Individual differences in the mean values between variants and criteria are thus preserved, while enabling the comparison of subjective ratings across the entire dataset, independently from individual scale usage. Interpreting the transformed values is, however, more difficult. Previous works have shown that different transformations can be beneficial where data is being prepared with other primary goals in mind, such as strictly normalizing the distribution of results to enable parametric  statistical analysis \cite{harrer2007characterisation}, or for an immediate interpretation of central tendencies without further statistical analysis \cite{gomez2015findings}. In the present case, the primary objective is to homogenize the variance distribution between drivers while retaining the resolution within a set of evaluations, which can be achieved by applying the z-transform, as shown in \cite{data2002objective,zschocke2008links}.

The ordinal data obtained from the second questionnaire are already in a symmetrical scale format with a minimum value of $-3$ and a maximum value of $3$. The sign convention is identical to that of the transformed BI. The ordinal data show considerably more homogenous scale usage across all user levels. Due to the limited power of parametric statistical analysis for ordinal data, these data are not normalized before statistical analysis. Data preparation was performed using MATLAB R2023a. Statistical analysis was implemented in jamovi version 2.3.28 \cite{jamovi2024jamovi}.
\section{Statistical analysis}
An analysis of variances (ANOVA) was performed on the entire dataset of transformed BI-ratings collected in the first questionnaire to investigate the main effects on the subjective evaluation of steering feedback with a significance level of~$.05$. All data were tested for normality using a Shapiro-Wilk test and for homogeneity of variances using Levene's method. A conventional parametric repeated-measures ANOVA was performed for normally distributed subjective data. Pairs of variants were compared using Tukey's HSD test. For the majority of subjective data that did not have a normal distribution or showed significant violations of homogeneity of variance, analysis of variances was performed by means of a non-parametric Friedman's test. Pairs of variants were then compared using jamovi's Durbin-Conover method \cite{pohlert2014pairwise} (PMCMR). All comparison ratings from the second questionnaire were analyzed through a Friedman test.

\subsection{BI ratings of simulator variants and reference vehicle}
A repeated-measures ANOVA revealed a highly significant difference in subjective ratings of the parameter Road contact \anovaPL{4}{120}{8.37}{} with $\eta_p^2 = .218$. A Friedman test revealed a significant difference in subjective ratings of the parameters SWT gradient \ChiSquare{4}{14.8}{.005}, LF Steering feedback \ChiSquare{4}{10.3}{.036}, HF Steering feedback \ChiSquarePL{4}{41.2}{}, LF Chassis feedback \ChiSquare{4}{12.5}{.014}, and HF Chassis feedback \ChiSquarePL{4}{22.1}{}. Pairwise comparisons of these parameters between variants that exhibit significant differences are presented in \autoref{tab_pairwise_BI}. All other questions did not show significant differences between either the reference vehicle and simulator variants or between simulator variants.

\subsection{Comparison ratings of simulator variants}
A Friedman test revealed a highly significant difference in subjective ratings of the parameter HF Chassis feedback \ChiSquare{3}{11.5}{.009} between at least two simulator variants. Pairwise comparisons of this parameter are presented in \autoref{tab_pairwise_Comp}.

\begin{table}[]
    \label{tab_pairwise_BI}
    \centering
    \caption{Pairwise comparison of BI ratings. The variant that performed better is shown in bold. Positive mean difference means B was rated higher/stronger/steeper than A.}
    \begin{tabular}{lllrrrr}
    \toprule
    \multicolumn{3}{c}{\textbf{Comparison}}     & \multicolumn{2}{c}{\textbf{Mean difference}}      & \multicolumn{2}{c}{\textbf{p}}    \\[0.05cm] 
    A                   & - & B                 &       Overall     &       Experts         &       Overall     &       Experts                 \\      \midrule
    \multicolumn{6}{l}{\textit{SWT gradient} \textsuperscript{**}}                                              \\      \midrule
    \textbf{Ref}        & - & V2                &  $ 0.642$     &  $ 0.577$     & $ .007$      & $ .021$      \\
                        & - & V3                &  $ 0.542$     &  $ 0.637$     & $ .054$      & $ .010$      \\      \midrule
    \multicolumn{6}{l}{\textit{Road contact} \textsuperscript{*}}                                             \\      \midrule
    \textbf{Ref}        & - & V1                & $-0.279$      & $-0.395$      & $ .008$      & $ .004$      \\
                        & - & V2                & $-0.447$      & $-0.616$      & $<.001$      & $<.001$      \\          
                        & - & V4                & $-0.466$      & $-0.556$      & $ .002$      & $ .003$      \\
        V2              & - & \textbf{V3}       & $ 0.218$      & $ 0.396$      & $ .087$      & $ .003$      \\ 
    \textbf{V3}         & - & V4                & $-0.238$      & $-0.336$      & $ .141$      & $ .049$      \\      \midrule
    \multicolumn{6}{l}{\textit{Low-frequency steering feedback} \textsuperscript{**}}                          \\      \midrule
    \textbf{Ref}        & - & V2                & $ 0.264$      & $ 0.574$      & $ .040$      & $ .017$      \\
                        & - & V4                & $ 0.037$      & $ 0.358$      & $ .068$      & $ .016$      \\      \midrule
    \multicolumn{6}{l}{\textit{High-frequency steering feedback} \textsuperscript{**}}                         \\      \midrule
    \textbf{Ref}        & - & V1                & $ 0.530$      & $ 0.726$      & $<.001$      & $<.001$      \\
                        & - & V2                & $ 0.829$      & $ 0.894$      & $<.001$      & $<.001$      \\          
                        & - & V4                & $ 0.857$      & $ 0.941$      & $<.001$      & $<.001$      \\
        V2              & - & \textbf{V3}       & $-0.518$      & $-0.541$      & $ .001$      & $ .001$      \\
    \textbf{V3}         & - & V4                & $ 0.546$      & $ 0.589$      & $<.001$      & $<.001$      \\      \midrule
    \multicolumn{6}{l}{\textit{Low-frequency chassis feedback} \textsuperscript{**}}                          \\      \midrule
    \textbf{Ref}        & - & V2                & $ 0.500$      & $ 0.878$      & $ .025$      & $ .007$      \\  
    \textbf{V1}         & - & V2                & $ 0.487$      & $ 0.518$      & $ .044$      & $ .064$      \\ 
        V2              & - & \textbf{V3}       & $-0.385$      & $-0.529$      & $ .044$      & $ .055$      \\ 
                        & - & \textbf{V4}       & $-0.522$      & $-0.735$      & $ .044$      & $ .011$      \\      \midrule
    \multicolumn{6}{l}{\textit{High-frequency chassis feedback} \textsuperscript{**}}                          \\      \midrule
    \textbf{Ref}        & - & V1                & $ 0.253$      & $ 0.592$      & $ .129$      & $ .014$      \\ 
                        & - & V2                & $ 0.740$      & $ 1.008$      & $<.001$      & $<.001$      \\  
                        & - & V4                & $ 0.380$      & $ 0.808$      & $ .026$      & $ .003$      \\
        V2              & - & \textbf{V3}       & $-0.429$      & $-0.515$      & $ .007$      & $ .006$      \\
    \bottomrule
    \vspace{-0.2cm} \\
    \multicolumn{6}{l}{\textsuperscript{~ *}: Tukey's HSD test} \\
    \multicolumn{6}{l}{\textsuperscript{**}: PMCMR with Holm-Bonferroni correction} \\
    \end{tabular}
    \end{table}

\begin{table}[]
    \centering
    \caption{Pairwise comparison of comparison ratings (PMCMR with Holm-Bonferroni correction). The variant that performed better is shown in bold.}
    \label{tab_pairwise_Comp}
    \begin{tabular}{lllrrrr}
    \toprule
    \multicolumn{3}{c}{\textbf{Comparison}} & \multicolumn{2}{c}{\textbf{Mean difference}}      & \multicolumn{2}{c}{\textbf{p}}    \\[0.05cm] 
    A                   & - & B             & Overall               & Experts                   & Overall   & Experts               \\ \midrule
    \multicolumn{5}{l}{\textit{High-frequency chassis feedback}}                                                                    \\ \midrule
            V2          & - & \textbf{V3}               & $ 0.844$  & $ 0.767$                  & $ .007$  & $ .010$              \\
        \textbf{V3}     & - &   V4                      & $-0.570$  & $-0.578$                  & $ .047$  & $ .077$              \\ \bottomrule
    \end{tabular}
    \end{table}

\subsection{Main effects of simulator variations}
To test how the independent variables \enquote{Tire Model} (MFTire or FTire) and \enquote{SST Status} (SSTs on or off) affected the subjective evaluation of steering feedback, a mixed-factorial ANOVA with the between-subjects factor \enquote{User Level} was performed for parameters following a normal distribution. Parameters not following a normal distribution were analyzed with a non-parametric Wilcoxon rank test.
In the first questionnaire, a repeated-measures ANOVA showed that the main effect of Tire Model on the subjective evaluation of Road contact was not significant but there was a significant main effect of SST Status. A Wilcoxon rank test showed that SST Status had a highly significant effect on subjective evaluation of HF Steering feedback (\wilcoxonPL{73.0}{}) and a significant effect on LF Chassis feedback (\wilcoxon{112}{.023}) and HF Chassis feedback (\wilcoxon{145}{.045}). Tire model had no significant effect on the evaluation criteria across the entire dataset. 

In the second questionnaire, a Wilcoxon rank test showed that SST Status had a significant effect on the subjective evaluation of comparison ratings of HF Chassis feedback (\wilcoxon{129}{.019}). Tire Model had no significant effect on the evaluation criteria across the entire dataset.

\subsection{Differences between user levels}
To check for interaction effects between the variants and the expertise level of the users, all data were additionally investigated for significance regarding the effect of the parameter User Level on mean value differences. A repeated-measures ANOVA revealed a significant interaction effect of User Level on Tire Model in the subjective ratings of the parameter Road contact \anova{2}{30}{5.820}{.007} with $\eta_p^2 = .280$. This effect is visualized through an analysis of marginal means in \autoref{fig_MargMeans_UserLevel}.
While the majority of participants had extensive experience on the driving simulator, the subgroup of participants with User Level $0$ contained a small number of drivers without any previous simulator experience. This provides an explanation for the considerably less conclusive findings from the entire dataset which was observed in the analysis of main effects. To prevent the negative effect of less experienced drivers on the dataset, the overall number of participants was deliberately increased beyond the minimum sample size to ensure that a subgroup analysis would provide sufficient statistical power. Therefore, the presented subgroup analysis of experts needs to be considered since it provides a more representative sample of the targeted participants. 

\subsubsection{Subgroup analysis of BI ratings of experts}
In the subgroup experts, a repeated-measures ANOVA showed a highly significant difference in subjective ratings of the parameter Road contact \anovaPL{4}{92}{10.3}{} with $\eta_p^2 = .309$. A Friedman test revealed a significant difference in subjective ratings of the parameters SWT gradient \ChiSquare{4}{13.3}{.01}, LF Steering feedback \ChiSquare{4}{13.0}{.011}, HF Steering feedback \ChiSquarePL{4}{36.9}{}, LF Chassis feedback \ChiSquare{4}{14.4}{.006}, and HF Chassis feedback \ChiSquarePL{4}{24.9}{}. Pairwise comparisons of these parameters between variants that exhibit significant differences are presented in \autoref{tab_pairwise_BI}.

\subsubsection{Subgroup analysis of comparison ratings of experts}
In the subgroup experts, a Friedman test showed a highly significant difference in subjective ratings of the parameter HF Chassis feedback \ChiSquare{3}{10.5}{.015}. Pairwise comparisons of this parameter are presented in \autoref{tab_pairwise_Comp}.

\section{Discussion}
The results of the statistical analysis of z-transformed subjective data confirmed the conclusions from the ranking of raw data and were consistent with the overall subjective feedback of participants. Variant V3 was rated best and most comparable to the reference vehicle. BI ratings showed considerably lower mean value differences between the reference vehicle and V3 compared to the other simulator variants across all user groups. These findings align with the rankings from the final post-study evaluation of overall steering feedback realism, in which eleven participants chose variant V3, ten participants chose V1, and both V2 and V4 were chosen by six participants. With the exception of the parameter SWT gradient, subjective data showed statistically significant differences only in the characteristics related to Road contact and Chassis feedback. Pairwise comparisons showed that significant differences in the parameters SWT gradient and LF Steering feedback occur between the reference vehicle and simulator variants, but not between simulator variants. In both parameters, the entire dataset showed statistically significant differences only between the reference vehicle and variant V2, while the analysis of the subgroup experts revealed another significant difference. All other parameters showed significant differences between variant V3 and at least one other simulator variant. Variant V3 is the only simulator variant that did not show significant differences to the reference vehicle in any parameter. 

\begin{figure*}[!t]
    \centering
    \subfloat[]{\includegraphics[width=3.5in]{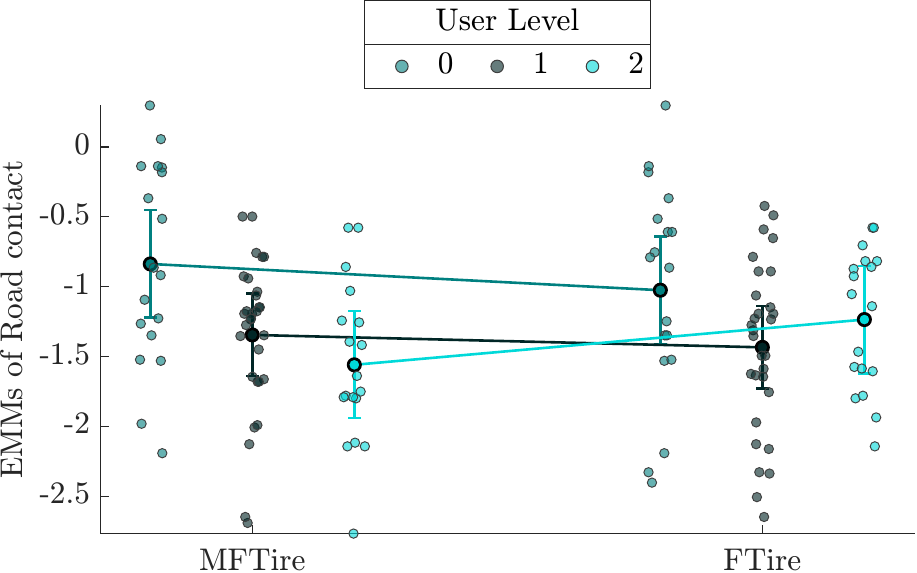}%
    \label{fig_MargMeans_TireModelUserLevel}}
    \hfil
    \subfloat[]{\includegraphics[width=3.5in]{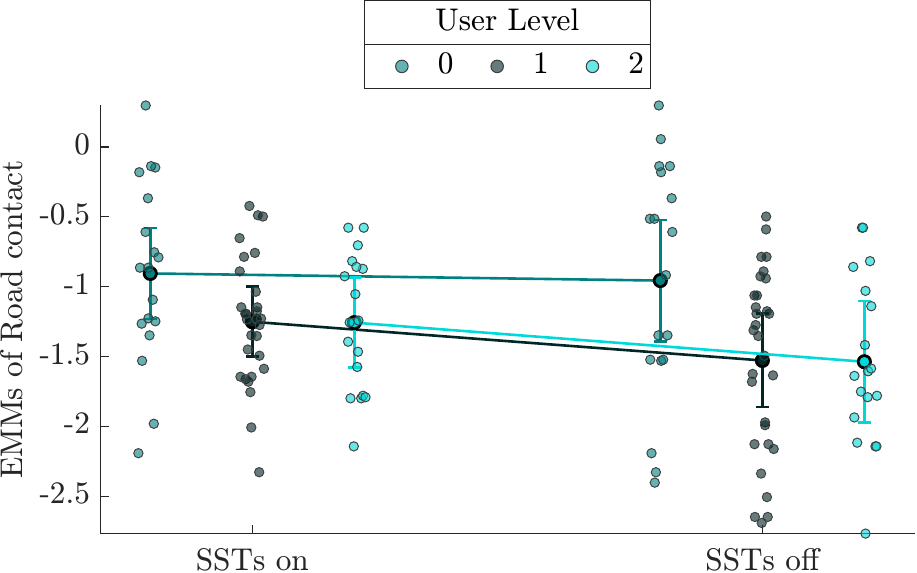}%
    \label{fig_MargMeans_KSWUserLevel}}
    \caption{Estimated marginal means (EMMs) and $95\%$ confidence intervals of (a) Tire model and (b) SST status separated by User Level}
    \label{fig_MargMeans_UserLevel}
\end{figure*}

In this study, variants providing a higher level of both steering and chassis feedback were rated better and closer to the reference vehicle in all criteria demonstrating significant differences between variants, as can be observed in \autoref{tab_pairwise_BI}. In all statistically significant pairwise comparisons, variants with the FTire model performed better than variants with the MFTire model, and variants with active SSTs performed better than variants with passive SSTs. The subgroup experts showed significant differences in the same parameters as the overall analysis, but in most cases with higher effect sizes. Due to the large number of participants with User Levels $1$ and $2$, the post-hoc subgroup analysis of experts in \autoref{tab_pairwise_BI} provides additional insights, with sufficient statistical power of $.873$ for medium sized effects with \anova{4}{92}{2.471}{.05}. While there was no significant main effect of Tire model across all participants, the marginal means in \autoref{fig_MargMeans_TireModelUserLevel} show a large interaction effect of User Level and Tire model with $\eta_p^2 = .280$. While participants with User Levels $0$ and $1$ rated the variants with the MFTire model with moderately higher scores, the participants with User Level $2$ rated the variants with the FTire model with considerably higher scores. This difference in effect directions provides an explanation for the inconclusive main effect observed across the entire dataset. As illustrated by the marginal means in \autoref{fig_MargMeans_KSWUserLevel}, the main effect of SST status with $\eta_p^2 = .266$ can be observed irrespective of User Level. Participants with User Level $0$ rated variants with active SSTs with moderately higher scores while participants with User Levels $1$ and $2$ rated the variants with active SSTs with considerably higher scores. 
In summary, the statistical analysis of subjective data confirmed the contribution of non-road-induced excitations to the subjective evaluation of steering feedback, with a clear preference for variants with active SSTs. While a subgroup analysis further revealed Tire Model as a contributory factor, and post-hoc analysis indicated a preference for variants with the FTire model, the results were less conclusive with regard to the contribution of road-induced excitations.

\section{Limitations}
\label{chap_Limitations}
Due to the imbalanced gender distribution among the participants, the results of this study are not representative of the general population. Since the study design required a highly specific evaluation task tailored towards the specifically trained and experienced drivers that are typically involved in comparable driving simulator evaluations in the course of the development process, recruiting a sufficient number of participants with an adequate level of expertise and simulator experience was crucial. Since an exploration of age- or gender-specific differences might provide additional insights beyond the scope of this work, a similar study with a more balanced sample of participants could be considered in future research.

Furthermore, a conceptual limitation exists with regard to the comparability of closed-loop experimental data through training effects and the limited repeatability of driver behavior in consecutive evaluation drives. In this study, this effect was mitigated by the use of the drivers' own velocity profiles in the simulator drives. While this does provide objective comparability of the evaluated variants and eliminates the effect of driving speed on the subjective evaluation, the resulting differences in driver workload between the reference vehicle and the simulator variants remain a source of uncertainty, the effect of which cannot be isolated in this study design. In the presented study, participants were asked to use the reference vehicle's cruise control system during the reference drive to minimize these differences. Nevertheless, an investigation of the effects on driver workload and performance in similar, comparative closed-loop simulator studies is recommended for future research. 

Lastly, the dynamic system limitations of the driving simulator used in this study regarding the representation of both rotational steering wheel accelerations and chassis excitations restrict the effect of the differences in road-induced steering wheel and chassis vibrations between the different tire models. This provides an explanation for the small effects that were observed with respect to this modification. An investigation utilizing a simulator system with higher dynamic frequency limitations in future research could therefore provide a clearer picture of the contribution made by different tire models to the subjective evaluation of steering feedback.

\section{Conclusion}
In summary, the effects of road-induced and non-road-induced steering wheel and chassis vibrations in the frequency range between $30$ and \SI{100}{\hertz} on the subjective evaluation of steering feedback were investigated in a controlled back-to-back study, comparing a reference vehicle and four variants of its virtual representation in a dynamic high-fidelity driving simulator. Four simulator variants were evaluated in a single-blind within-subjects design in direct comparison after a drive in the reference vehicle. Two different states of road-induced excitations applied through the motion platform were implemented via one semi-empirical and one physics-based tire model, and two different states of non-road-induced excitations were implemented through model-based high-frequency excitations that were applied through SSTs. 

Results showed that variant V3 -- using a physics-based tire model in combination with non-road-induced excitations through SSTs -- performed best and was the only variant that did not show significant differences to the reference vehicle. Across all simulator variants, the addition of high-frequency non-road-induced vibrations showed a clear benefit for the subjective evaluation of steering and chassis feedback, while the results were less conclusive regarding the beneficial effect of additional road-induced vibrations through a physics-based tire model. 

It is recommended that future research investigate the contribution of system excitations that were not within the scope of the presented study, such as higher excitation frequencies via the acoustic channel. 

\section*{Contribution}
Maximilian Böhle initiated the idea of this paper, and designed and implemented the presented subject study. Steffen Müller and Bernhard Schick made essential contributions to the conception of the research project and the study design, and critically revised the paper.

\section*{Acknowledgments}
Special thanks go to Mike Koehler and Philipp Rupp at Kempten University of Applied Sciences for their support during the operation of the driving simulator at the IFM.

\bibliographystyle{IEEEtran}
\bibliography{IEEEabrv,2025_IEEE_T-IV.bib}
\FloatBarrier
\vspace{-5cm}

\begin{IEEEbiography}[{\includegraphics[width=1in,height=1.25in,clip,keepaspectratio]{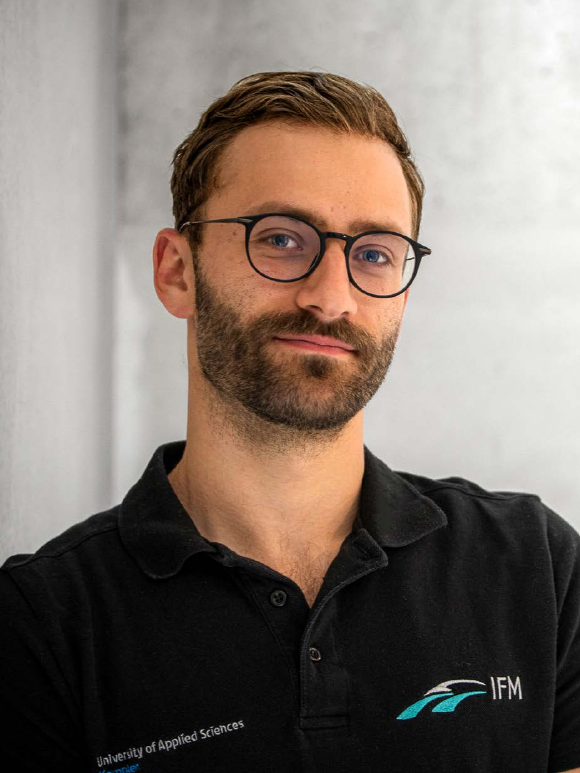}}]{Maximilian Böhle}
  was awarded B.Sc. and M.Sc. degrees in Automotive Engineering by Munich University of Applied Sciences, Germany in 2016 and 2019. He is currently working towards his doctoral degree in Automotive Engineering at the Faculty of Mechanical Engineering and Transport Systems of the Technical University of Berlin, Germany. Since 2020, he has been working as a research assistant at the Institute for Driver Assistance and Connected Mobility at Kempten University of Applied Sciences, Germany. His research focuses on steering feel and vehicle dynamics for driving simulators.
\end{IEEEbiography}
\vspace{-5cm}
\begin{IEEEbiography}[{\includegraphics[width=1in,height=1.25in,clip,keepaspectratio]{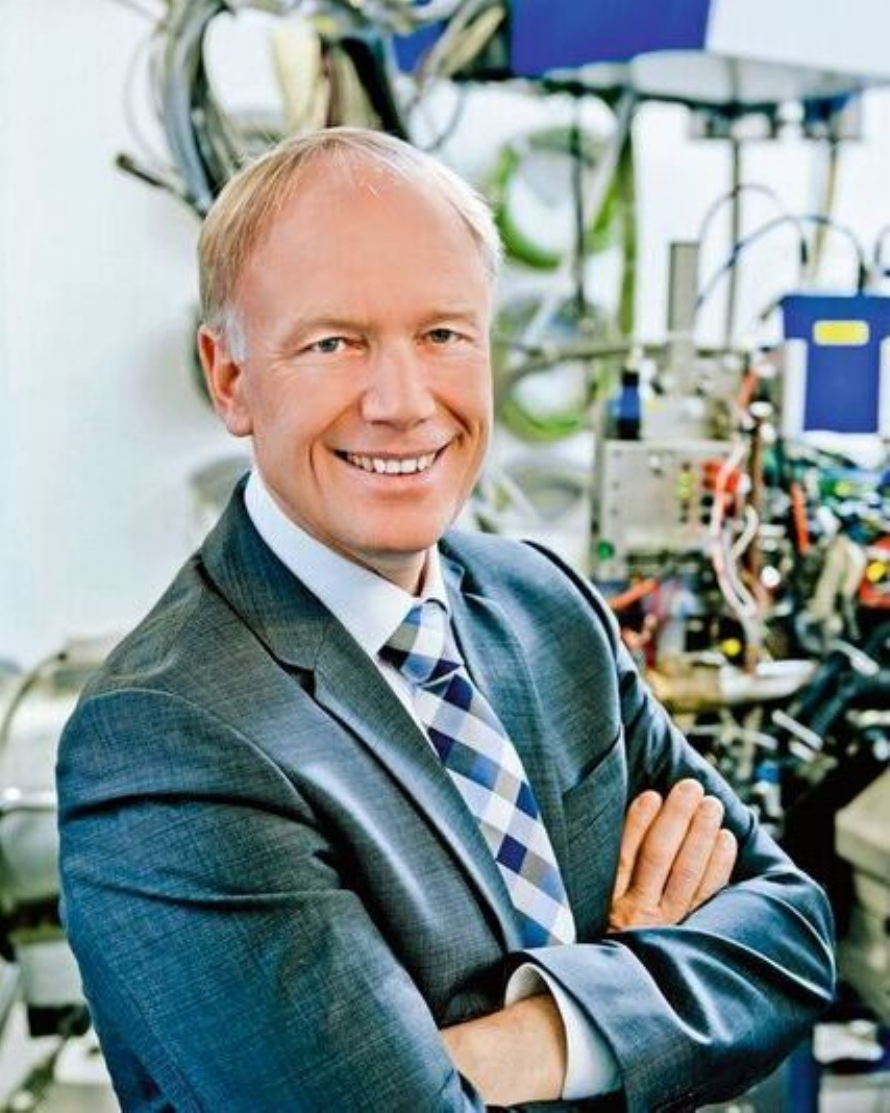}}]{Bernhard Schick}
  holds a degree in Mechatronic Engineering at the University of Applied Sciences Heilbronn. From 1994, whilst at TÜV SÜD, he honed his expertise in the field of vehicle dynamics and advanced driver assistance systems, in various positions up to general manager. He joined IPG Automotive in 2007 as managing director, where he worked in the field of vehicle dynamics simulation. From 2014, he was responsible for calibration and virtual testing technologies as global business unit manager at AVL List, Graz in Austria. Since 2016, he has been a research professor at Kempten University of Applied Sciences and the Head of the Institute for Driver Assistance Systems and Connected Mobility. His research focus is automated driving and vehicle dynamics.
\end{IEEEbiography}
\vspace{-5cm}
\begin{IEEEbiography}[{\includegraphics[width=1in,height=1.25in,clip,keepaspectratio]{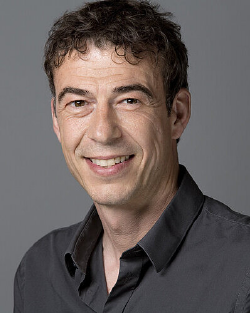}}]{Steffen Müller}
  was awarded his Dipl.-Ing. degree in Astronautics and Aerospace Engineering in 1993 and Dr.-Ing. degree by the Technical University of Berlin in 1998. From 1998 to 2000, he was a project manager at the ABB Corporate Research Center, Heidelberg, Germany. He finished his post-doctoral research at the University of California, Berkeley, in 2001. From 2001 to 2008, he held leading positions at the BMW Research and Innovation Center. From 2008 to 2013, he was the founder and leader of the Chair for Mechatronics in Engineering and Vehicle Technology, Technical University of Kaiserslautern, Germany. He is a university professor and an Einstein Professor at the Technical University of Berlin, Germany. He is the Head of the Chair of Automotive Engineering in the Faculty of Mechanical Engineering and Transport Systems there.
\end{IEEEbiography}

\end{document}